\documentclass[%
 reprint,
showpacs,
preprintnumbers,
 amsmath,
 amssymb,
 aps,
 prd,
]{revtex4-1}

\usepackage{graphicx}
\usepackage{dcolumn}
\usepackage{bm}

\begin{document}

\title{A one-dimensional soliton system of gauged Q-ball and anti-Q-ball}
\author{A. Yu. Loginov}
\email{aloginov@tpu.ru}
\author{V. V. Gauzshtein}
\affiliation{Tomsk Polytechnic University, 634050 Tomsk, Russia}

\date{\today}

\begin{abstract}
The $(1+1)$-dimensional  gauge  model  of  two  complex self-interacting scalar
fields that  interact  with  each  other  through  an Abelian gauge field and a
quartic scalar interaction is considered.
It is shown that the model  has  nontopological  soliton  solutions  describing
soliton  systems  consisting  of  two  Q-ball  components  possessing  opposite
electric charges.
The  two  Q-ball  components interact with each other through the Abelian gauge
field and the quartic scalar interaction.
The interplay  between  the  attractive  electromagnetic  interaction  and  the
repulsive   quartic  interaction  leads  to  the  existence  of  symmetric  and
nonsymmetric soliton systems.
Properties  of  these  systems  are  investigated  by  analytical and numerical
methods.
The symmetric  soliton  system  exists  in  the whole allowable interval of the
phase frequency, whereas the  nonsymmetric  soliton  system exists only in some
interior subinterval.
Despite the  fact  that  these  soliton  systems are electrically neutral, they
nevertheless possess nonzero electric fields in their interiors.
It is found that the nonsymmetric  soliton  system  is more preferable from the
viewpoint of energy than the symmetric one.
Both symmetric and  nonsymmetric  soliton  systems are stable to the decay into
massive scalar bosons.
\end{abstract}

\maketitle

\section{\label{sec:I} Introduction}

There are many  field  models  possessing  global  symmetries and corresponding
conserved Noether charges that admit the existence  of  nontopological solitons
\cite{lee, radu}.
The determining property of  a nontopological soliton is that it is an extremum
of the energy functional at a fixed value of the Noether charge.
This feature  of  nontopological  solitons  leads  to  the  characteristic time
dependence $\propto \exp \left( - i\omega t\right)$  of their fields.
This nontrival time dependence of the  soliton's  field  allows to avoid severe
restrictions of  Derrick's  theorem  \cite{derrick},  so  scalar nontopological
solitons can exist in any number of spatial dimensions.

The simplest  nontopological  soliton,  proposed  in \cite{rosen} and  known as
a Q-ball \cite{coleman},  has been found  in  a model of a complex scalar field
possessing a global $U\left(1\right)$ symmetry.
Q-balls can  also  exist in scalar field models possessing a global non-Abelian
symmetry \cite{saf1, saf2}.
They  are  present \cite{kus2, kst}  in  the  minimal  supersymmetric extension
of the Standard Model  having  flat  directions in the interaction potential of
scalar fields.
Q-balls are of great interest to cosmological models  describing  the evolution
of the early Universe \cite{kus3, enq}.

There are also other types of nontopological solitons in global-symmetric field
models.
The most known of them is the nontopological soliton of the Friedberg-Lee-Sirlin
model \cite{fried}.
The model consists of two interacting  scalar  fields, one of which is real and
the other is complex.
It  possesses  a  global  $U\left( 1 \right)$  symmetry  and  a  renormalizable
interaction potential.
Another example  is  the  nontopological  soliton  in  the  model  of a massive
self-interacting complex vector field \cite{loginov_2015}.

In all of the examples given above, the existence of nontopological solitons is
due to a global invariance  of  the  corresponding Lagrangians, so the  Noether
charge of such solitons cannot be a source of a gauge field.
At the same time, nontopological solitons also exist in field models possessing
a local  gauge  invariance, both Abelian \cite{klee, lee_yoon_1991, anag, levi,
ardoz_2009, gulamov_2014} and non-Abelian \cite{fried1, fried2}.
The  nontopological solitons \cite{klee, lee_yoon_1991, anag, levi, ardoz_2009,
 gulamov_2014} possess a long-range  gauge  electric field, and Noether charges
of these solitons are proportional to their electric charges.
However,   all   these   electrically   charged   nontopological  solitons  are
three-dimensional ones.
This is  because  any  one-dimensional  or two-dimensional  field configuration
with a nonzero electric charge  possesses  infinite  energy, as it follows from
Gauss's  law  and  the  expression  for  the electric field energy density.
Nevertheless, there are  electrically  neutral  low-dimensional soliton systems
that have a nonzero electric field in their interiors.
In particular,  the  two-dimensional  soliton systems  consisting of vortex and
Q-ball  components  interacting  through  an  Abelian  gauge  field  have  been
described in \cite{loginov_plb_777, loginov_plb_784}.

In the present  paper,  we  research the $(1+1)$-dimensional gauge model of two
complex self-interacting scalar  fields  interacting with each other through an
Abelian gauge field and a quartic scalar interaction.
In  particular, it  is  found  that  symmetric and nonsymmetric soliton systems
exist in the model.
The soliton systems consist of two Q-ball  components  having opposite electric
charges.
The soliton systems are electrically neutral  but  nevertheless possess nonzero
electric fields in their interiors.
The  paper  is  structured as follows.
In Sec.~\ref{sec:II}, we describe briefly the Lagrangian and the field equations
of the model under consideration.
By means of the Hamiltonian formalism  and the Lagrange multipliers method, the
time dependence is established for the soliton system's fields.
Then, we  give  the  ansatz  used  for  solving the model's field equations and
establish the basic relation for the nontopological soliton system.
In Sec.~\ref{sec:III}, we derive the system of nonlinear differential equations
for the  ansatz  functions  and the expressions for the electromagnetic current
density and the energy density in terms of these functions.
Then, some  general  properties  of  the  soliton  system  are established, its
asymptotic properties are researched, and the virial relation  for  the soliton
system is derived.
In Sec.~\ref{sec:IV},  we  study  properties  of  the  soliton  system  in  the
thick-wall and thin-wall regimes and establish its stability to decay into free
massive scalar bosons.
In Sec.~\ref{sec:V}, we  briefly describe  the  procedure for numerical solving
of a boundary value  problem  and  discuss  possible types of soliton solutions
of the problem.
The dependences  of  the  energy  and the Noether charge on the phase frequency
are  presented  for  both (symmetric  and  nonsymmetric)  types  of the soliton
solutions.
Then, we show the dependences  of  the  symmetric  soliton  system's energy and
the energy difference between the symmetric and  nonsymmetric  soliton  systems
on the Noether charge.
After that,  we  present  the  numerical  results for the ansatz functions, the
energy density, the  electric  charge  density, and the electric field strength
for  the symmetric and nonsymmetric soliton  systems.

Throughout the paper the natural units $c = 1$, $\hbar = 1$ are used.

\section{\label{sec:II} The Lagrangian and the field equations}

The $(1 + 1)$-dimensional gauge model  we are interested in is described by the
Lagrangian density
\begin{eqnarray}
\mathcal{L} &=&-\frac{1}{4}F_{\mu \nu }F^{\mu \nu }+\left( D_{\mu }\phi
\right) ^{\ast }D^{\mu }\phi +\left( D_{\mu }\chi \right) ^{\ast }D^{\mu
}\chi \nonumber \\
&&-V\left( \left\vert \phi \right\vert \right) -U\left( \left\vert \chi
\right\vert \right) -W\left( \left\vert \phi \right\vert ,\left\vert \chi
\right\vert \right).                                               \label{II:1}
\end{eqnarray}
It describes  the  two  complex  scalar fields $\phi$ and $\chi$ that minimally
interact  with  the  Abelian   gauge  field  $A_{\mu}$  through  the  covariant
derivatives:
\begin{equation}
D_{\mu }\phi  = \partial _{\mu }\phi + ieA_{\mu }\phi,\quad
D_{\mu }\chi  = \partial _{\mu }\chi + iqA_{\mu }\chi.             \label{II:2}
\end{equation}
The scalar fields interact with each other and self-interact.
The  self-interaction  potentials   of   the   scalar  fields   have  the  form
\begin{eqnarray}
V\left( \left\vert \phi \right\vert \right) & = & m_{\phi }^{2}\left\vert \phi
\right\vert ^{2}-\frac{g_{\phi }}{2}\left\vert \phi \right\vert ^{4~} + \frac{
h_{\phi }}{3}\left\vert \phi \right\vert ^{6},                    \label{II:3a}
\\
U\left( \left\vert \chi \right\vert \right) & = & m_{\chi }^{2}\left\vert \chi
\right\vert ^{2}-\frac{g_{\chi }}{2}\left\vert \chi \right\vert ^{4~} + \frac{
h_{\chi }}{3}\left\vert \chi \right\vert ^{6},                    \label{II:3b}
\end{eqnarray}
whereas the interaction potential is
\begin{eqnarray}
W\left(\left\vert \phi \right\vert, \left\vert \chi \right\vert \right)
&=&\lambda \left\vert
\phi \right\vert ^{2}\left\vert \chi \right\vert^{2}.              \label{II:4}
\end{eqnarray}
We suppose that the self-interaction potentials $V$ and $U$ admit the existence
of usual non-gauged  nontopological  solitons (Q-balls) formed  from the scalar
fields $\phi$ and $\chi$, respectively.
We also suppose that the  potentials $V$ and $U$ possess global minima at $\phi
= 0$ and $\chi = 0$, respectively.
Then the parameters of the potentials satisfy the condition
\begin{equation}
\frac{m_{i}^{2} h_{i}}{g_{i}^{2}} > \frac{3}{16},                  \label{II:5}
\end{equation}
where the index $i = \left( \phi, \chi \right)$.

The Lagrangian (\ref{II:1}) is invariant under the local gauge transformations.
At the same time, it is also invariant  under  the two independent global gauge
transformations:
\begin{subequations} \label{II:6}
\begin{eqnarray}
\phi \left( x\right)  &\rightarrow &\phi ^{\prime }\left( x\right) = \exp
\left(-i \alpha \right) \phi \left( x\right),
\\
\chi \left( x\right)  &\rightarrow &\chi ^{\prime }\left( x\right) = \exp
\left(-i \beta \right) \chi \left( x\right).
\end{eqnarray}
\end{subequations}
The  Noether currents corresponding to transformations (\ref{II:6}) are written
as
\begin{subequations} \label{II:7}
\begin{eqnarray}
j_{\phi }^{\mu } &=&i\left[ \phi ^{\ast }D^{\mu }\phi - \left(D^{\mu }\phi
\right) ^{\ast }\phi \right],
\\
j_{\chi }^{\mu } &=&i\left[ \chi ^{\ast }D^{\mu }\chi -\left( D^{\mu }\chi
\right) ^{\ast }\chi \right].
\end{eqnarray}
\end{subequations}
The presence  of  the  two  separately  conserved Noether charges  $Q_{\phi } =
\int j_{\phi }^{0}dx$ and $Q_{\chi } = \int j_{\chi }^{0}dx$  is  the result of
the structure  of  the  interaction  potential  $W$  and  the neutrality of the
Abelian gauge field $A_{\mu}$.

The field equation of the model are obtained  by  varying  the action $S = \int
\mathcal{L}d^{2}x$ in the corresponding fields:
\begin{eqnarray}
D_{\mu }D^{\mu }\phi +\frac{\partial V}{\partial \left\vert \phi \right\vert
}\frac{\phi }{2\left\vert \phi \right\vert }+\frac{\partial W}{\partial
\left\vert \phi \right\vert }\frac{\phi }{2\left\vert \phi \right\vert } &=&0,
\label{II:8a} \\
D_{\mu }D^{\mu }\chi +\frac{\partial U}{\partial \left\vert \chi \right\vert
}\frac{\chi }{2\left\vert \chi \right\vert }+\frac{\partial W}{\partial
\left\vert \chi \right\vert }\frac{\chi }{2\left\vert \chi \right\vert } &=&0,
\label{II:8b} \\
\partial _{\mu }F^{\mu \nu } & = &j^{\nu },                       \label{II:8c}
\end{eqnarray}
where  the  electromagnet  current $j^{\nu}$ is written in terms of two Noether
currents (\ref{II:7})
\begin{equation}
j^{\nu }=ej_{\phi }^{\nu }+qj_{\chi }^{\nu }.                      \label{II:9}
\end{equation}
The symmetric energy-momentum tensor of the model is written as
\begin{eqnarray}
T_{\mu \nu }& =&-F_{\mu \lambda }F_{\nu }^{\;\lambda }+\frac{1}{4}g_{\mu \nu
}F_{\lambda \rho }F^{\lambda \rho } \nonumber \\
& &+\left( D_{\mu }\phi \right) ^{\ast }D_{\nu }\phi +\left( D_{\nu }\phi
\right) ^{\ast }D_{\mu }\phi \nonumber \\
& &+\left( D_{\mu }\chi \right) ^{\ast }D_{\nu }\chi +\left( D_{\nu }\chi
\right) ^{\ast }D_{\mu }\chi  \nonumber \\
& &-g_{\mu \nu }\left[ \left( D_{\mu }\phi \right) ^{\ast }D^{\mu }\phi
+\left( D_{\mu }\chi \right) ^{\ast }D^{\mu }\chi \right. \nonumber \\
& &\left. -V\left( \left\vert \phi \right\vert \right) -U\left( \left\vert
\chi \right\vert \right) -W\left( \left\vert \phi \right\vert ,\left\vert
\chi \right\vert \right) \right],                                 \label{II:10}
\end{eqnarray}
so we have the following expression for the energy density
\begin{eqnarray}
T_{00} = \mathcal{E} &=&\frac{1}{2}E_{x}^{2} +
\left( D_{t}\phi \right)^{\ast }D_{t}\phi
+\left( D_{x}\phi \right) ^{\ast }D_{x}\phi \nonumber \\
&&+\left( D_{t}\chi \right) ^{\ast }D_{t}\chi +\left( D_{x}\chi \right)
^{\ast }D_{x}\chi \nonumber \\
&&+V\left( \left\vert \phi \right\vert \right) +U\left( \left\vert \chi
\right\vert \right) +W\left( \left\vert \phi \right\vert ,\left\vert \chi
\right\vert \right).                                              \label{II:11}
\end{eqnarray}

By  analogy  with  nontopological   solitons,  we  find  a  solution  of  model
(\ref{II:1}) that is an extremum  of  the energy functional $E = \int \mathcal{
E} dx$ at a fixed value of the Noether charge $Q_{\chi} = \int j_{\chi}^{0}dx$.
Such a solution is an unconditional extremum of the functional
\begin{equation}
F = \int \mathcal{E} dx - \omega \int j_{\chi}^{0}dx =
E - \omega Q_{\chi},                                              \label{II:12}
\end{equation}
where $\omega$ is the Lagrange multiplier.
To determine  the  time  dependence  of  the  soliton solution, we will use the
Hamiltonian formalism.
We adopt the axial gauge in which  the spatial component of the gauge potential
vanishes: $A_{x} = A^{1} = 0$.
In this case, the gauge  model  is described in terms  of the eight canonically
conjugated fields:  $\phi$,  $\pi _{\phi } = \left( D_{0}\phi \right)^{\ast }$,
$\phi^{\ast}$, $\pi_{\phi^{\ast }} = D_{0}\phi$, $\chi$, $\pi _{\chi } = \left(
D_{0}\chi \right)^{\ast }$, $\chi^{\ast}$,  and $\pi_{\chi^{\ast }}=D_{0}\chi$.
Then, the Hamiltonian density has the form
\begin{eqnarray}
\mathcal{H}& = &\pi _{\phi }\partial _{t}\phi +\pi _{\phi ^{\ast }}\partial
_{t}\phi ^{\ast }+\pi _{\chi }\partial _{t}\chi +\pi _{\chi ^{\ast
}}\partial _{t}\chi ^{\ast }-\mathcal{L} \nonumber \\
& = & -\frac{1}{2}\left( \partial _{x}A_{0}\right) ^{2}+\pi_{\phi }\pi_{\phi
^{\ast }}+\pi _{\chi }\pi _{\chi ^{\ast }} \nonumber \\
& &  + \partial _{x}\phi ^{\ast }\partial _{x}\phi + \partial _{x}\chi^{\ast
}\partial _{x}\chi \nonumber \\
& & + ieA_{0}\left\{ \phi ^{\ast }\pi _{\phi ^{\ast }} - \phi \pi _{\phi
}\right\} + iqA_{0}\left\{ \chi ^{\ast }\pi _{\chi ^{\ast }}-\chi \pi _{\chi
}\right\} \nonumber \\
& & +V\left( \left\vert \phi \right\vert \right) +U\left( \left\vert \chi
\right\vert \right) +W\left( \left\vert \phi \right\vert ,\left\vert \chi
\right\vert \right),                                              \label{II:13}
\end{eqnarray}
where the  time  component  $A_{0}$  is  determined in terms of the canonically
conjugated fields by Gauss's law
\begin{equation}
\partial _{x}^{2}A_{0}+ie\left\{ \phi ^{\ast }\pi _{\phi ^{\ast }}-\phi \pi
_{\phi }\right\} + iq\left\{ \chi ^{\ast }\pi _{\chi ^{\ast }} - \chi \pi
_{\chi }\right\} = 0.                                             \label{II:14}
\end{equation}
Note that energy  density  (\ref{II:11})  does  not  coincide  with Hamiltonian
density (\ref{II:13}):
\begin{eqnarray}
\mathcal{H-E} &\mathcal{=}&\mathcal{-}\left( \partial _{x}A_{0}\right)
^{2}+ieA_{0}\left\{ \phi ^{\ast }\pi _{\phi ^{\ast }}-\phi \pi _{\phi
}\right\} \nonumber \\
&&+iqA_{0}\left\{ \chi ^{\ast }\pi _{\chi ^{\ast }}-\chi \pi _{\chi
}\right\}.                                                        \label{II:15}
\end{eqnarray}
However, the  integral  of  Eq.~(\ref{II:15}) over the space dimension vanishes
for field configurations possessing  finite  energy  and satisfying Gauss's law
(\ref{II:14}).
So, for such configurations
\begin{equation}
E = \int \mathcal{E} dx = H = \int \mathcal{H} dx.                \label{II:16}
\end{equation}

It can be shown that the field equations (\ref{II:8a}) and (\ref{II:8b}) can be
rewritten in the Hamiltonian form:
\begin{eqnarray}
\partial _{t}\phi  &=&\frac{\delta H}{\delta \pi _{\phi }}=\frac{\delta E}
{\delta \pi _{\phi }},\;\partial _{t}\pi _{\phi }=-\frac{\delta H}{\delta
\phi } = -\frac{\delta E}{\delta \phi },                         \label{II:17a}
\\
\partial _{t}\chi  &=&\frac{\delta H}{\delta \pi _{\chi }}=\frac{\delta E}
{\delta \pi _{\chi }},\;\partial _{t}\pi _{\chi }=-\frac{\delta H}{\delta
\chi } = -\frac{\delta E}{\delta \chi}.                          \label{II:17b}
\end{eqnarray}
Further, the first variation of  the  functional  $F$  vanishes for the soliton
solution:
\begin{equation}
\delta F = \delta E - \omega \delta Q_{\chi } = 0,                \label{II:18}
\end{equation}
where the  first variation of the Noether charge $Q_{\chi}$ can be expressed in
terms of the canonically conjugated fields
\begin{equation}
\delta Q_{\chi } = -i\int \left(\pi _{\chi}\delta \chi+\chi \delta \pi_{\chi}
-\text{c.c.}\right) dx.                                           \label{II:19}
\end{equation}
From  Eqs.~(\ref{II:17a}),  (\ref{II:17b}), (\ref{II:18}),  and  (\ref{II:19}),
we obtain  the following Hamilton field equations:
\begin{eqnarray}
\partial_{t}\chi  & = &\frac{\delta E}{\delta \pi_{\chi}} = \omega \frac{
\delta Q_{\chi }}{\delta \pi _{\chi }} = - i \omega \chi ,       \label{II:20a}
\\
\partial_{t}\chi^{\ast} & = & \frac{\delta E}{\delta \pi_{\chi^{\ast}}}
= \omega \frac{\delta Q_{\chi}}{\delta \pi_{\chi^{\ast}}} = i \omega \chi,
                                                                 \label{II:20b}
\end{eqnarray}
while  time  derivatives  of  the  other  model's  fields  are  equal  to zero.
Thus, in  the  adopted  gauge  $A_{x} = 0$, only  the  scalar  field $\chi$ has
nontrivial time dependence, whereas  the  model's  fields  $\phi$  and  $A_{0}$
do  not  depend  on  time:
\begin{subequations} \label{II:21}
\begin{eqnarray}
\phi \left( x,t\right)  & = & f\left( x \right),
\\
\chi \left( x,t\right)  & = & s\left( x \right) \exp\left(-i\omega t\right),
\\
A_{\mu}\left(x,t\right) & = & \left(a_{0}\left( x \right), 0 \right).
\end{eqnarray}
\end{subequations}
From  extremum  condition  (\ref{II:18}),  it follows that the soliton solution
satisfies the important relation
\begin{equation}
\frac{dE}{dQ_{\chi}} = \omega,                                    \label{II:22}
\end{equation}
where  the  Lagrange multiplier $\omega$ is some function of the Noether charge
$Q_{\chi}$.
Note that unlike Eqs.~(\ref{II:21}), relation (\ref{II:22}) is gauge-invariant.
Just  as in the case of non-gauged nontopological solitons \cite{lee}, relation
(\ref{II:22}) plays  the  primary  role  in  the  determining  of properties of
the gauged nontopological soliton system.

\section{\label{sec:III} Some properties of the solution}

In Eqs.~(\ref{II:21}), $f\left(x\right)$ and $s\left(x\right)$ are some complex
functions of the real argument $x$.
Substituting   Eqs.~(\ref{II:21})   into   field   equations   (\ref{II:8a}) --
(\ref{II:8c}), we obtain the system of ordinary nonlinear differential equations
for  the  functions  $a_{0}\left(x\right)$,  $f\left(x\right)$,  and  $s\left(x
\right)$:
\begin{eqnarray}
& & a_{0}^{\prime \prime }(x) - 2 \left( e^{2}\left\vert f\left( x\right)
\right\vert^{2}+q^{2}\left\vert s\left( x\right) \right\vert ^{2} \right)
a_{0}\left( x\right) \label{III:1} \\
& & + 2 q \omega \left\vert s\left( x\right) \right\vert ^{2} = 0, \nonumber
\end{eqnarray}
\begin{eqnarray}
& &f^{\prime \prime }\left( x\right) - \left( m_{\phi}^{2} -
e^{2} a_{0}\left( x\right) ^{2}\right) f\left( x\right)  \label{III:2} \\
& & +\left( g_{\phi }\left\vert f\left( x\right) \right\vert ^{2}-
h_{\phi}\left\vert f\left( x\right) \right\vert ^{4}-\lambda
\left\vert s\left(x\right) \right\vert ^{2}\right) f\left( x\right) = 0,
\nonumber
\end{eqnarray}
\begin{eqnarray}
& & s^{\prime \prime }\left( x\right) - \left( m_{\chi
}^{2} - \left(\omega - q a_{0}\left(x\right)\right)^{2}\right) s\left( x\right)
\label{III:3} \\
& & +\left( g_{\chi }\left\vert s\left( x\right) \right\vert ^{2}-h_{\chi
}\left\vert s\left( x\right) \right\vert ^{4}-\lambda \left\vert f\left(
x\right) \right\vert ^{2}\right) s\left( x\right) = 0. \nonumber
\end{eqnarray}
From  Eq.~(\ref{III:2}),  it  follows  that  the  real  and  imaginary parts of
$f\left( x \right)$   satisfy   the   same   differential   equation,   whereas
Eq.~(\ref{III:3})  leads   us   to   the  same   conclusion  for  the  function
$s\left(x\right)$.
This in turn means that  the  functions $f\left(x\right)$ and $s\left(x\right)$
can be written as $f\left( x \right) = \exp \left( i \alpha \right) \left \vert
f \left( x\right) \right \vert$ and $s\left(x\right)=\exp\left( i \beta \right)
\left\vert s\left( x \right) \right \vert$, where $\alpha$ and $\beta$ are real
constant phases.
These phases, however,  can  be  gauged  away  by  global gauge transformations
(\ref{II:6}).
Thus  we  can  suppose  without  loss  of generality that $f\left(x\right)$ and
$s\left(x\right)$ are real functions of $x$.
Substituting  Eqs.~(\ref{II:21})  into  Eq.~(\ref{II:9}) and Eq.~(\ref{II:11}),
we obtain the electromagnetic current density and  the  energy density in terms
of the real  functions $a_{0}\left(x\right)$, $f\left(x\right)$, and  $s\left(x
\right)$:
\begin{equation}
j^{\mu } = \left( 2q\omega s^{2} - 2\left( q^{2}s^{2} + e^{2}f^{2}\right)
a_{0},\,0\right),                                                 \label{III:4}
\end{equation}
\begin{eqnarray}
\mathcal{E} & = &\frac{a_{0}^{\prime }{}^{2}}{2}+f^{\prime
}{}^{2}+s^{\prime }{}^{2}+\left( \omega -qa_{0}\right)
^{2}s^{2}+e^{2}a_{0}{}^{2}f^{2} \nonumber \\
&&+V\left( f \right) +U\left( s \right) +W\left( f, s \right).    \label{III:5}
\end{eqnarray}
The finiteness of the soliton system's energy $E = \int \mathcal{E}dx$ leads to
the  following boundary  conditions  for  the  functions $a_{0}\left(x\right)$,
$f\left(x\right)$, and  $s\left(x\right)$:
\begin{subequations}\label{III:6}
\begin{eqnarray}
& a_{0}^{\prime }\left( x\right) \underset{x\rightarrow - \infty}
{\longrightarrow }0,\; &a_{0}^{\prime }\left( x\right)
\underset{x\rightarrow\infty }{\longrightarrow }0,
\label{III:6a} \\
& f\left( x\right) \underset{x\rightarrow -\infty }{\longrightarrow }
0,\; & f\left( x\right) \underset{x\rightarrow \infty }{\longrightarrow }0,
\label{III:6b} \\
& s\left( x\right) \underset{x\rightarrow -\infty }{\longrightarrow }
0,\; & s\left( x\right) \underset{x\rightarrow \infty }{\longrightarrow }0.
                                                                 \label{III:6c}
\end{eqnarray}
\end{subequations}

Let us discuss some general properties of the soliton system.
The  invariance  of  the  Lagrangian  (\ref{II:1}) under the charge conjugation
leads  to  the  invariance  of  system (\ref{III:1}) -- (\ref{III:3}) under the
discrete transformation
\begin{equation}
\omega, a_{0}, f, s \longrightarrow -\omega, -a_{0}, f, s.        \label{III:7}
\end{equation}
From Eqs.~(\ref{III:4}),  (\ref{III:5}), and (\ref{III:7}), it follows that the
energy  $E$  is  an  even  function  of  $\omega$,  whereas the Noether charges
$Q_{\phi}$ and $Q_{\chi}$ are odd functions of $\omega$:
\begin{eqnarray}
E\left( -\omega \right)  &=&E\left( \omega \right),
\label{III:8a} \\
Q_{\phi ,\chi }\left( -\omega \right)  &=&-Q_{\phi ,\chi }\left(\omega\right).
                                                                 \label{III:8b}
\end{eqnarray}

The Lagrangian (\ref{II:1}) is also invariant under the parity  transformation.
It follows  that  system (\ref{III:1}) -- (\ref{III:3})  is invariant under the
space inversion: $x \rightarrow -x$.
Thus, if $a_{0}\left(x\right)$, $f\left(x\right)$,  and  $s\left(x\right)$ is a
solution  of  Eqs.~(\ref{III:1}) -- (\ref{III:3}), then $a_{0}\left(-x\right)$,
$f\left(-x\right)$, and $s\left(-x\right)$ is also a solution.
This fact, however, does not mean that $a_{0}\left(x\right)$, $f\left(x\right)$,
and $s\left(x\right)$ must be even functions of $x$.
Indeed, we  shall see later that system (\ref{III:1}) -- (\ref{III:3}) together
with boundary  conditions  (\ref{III:6})  has  nonsymmetric  soliton solutions.

Eq.~(\ref{III:1}) can be  written  as $a_{0}^{ \prime \prime } = -j_{0}$, where
$j_{0}$ is electric charge density (\ref{III:4}).
Integrating this equation over $x \in \left(-\infty, \infty \right)$ and taking
into  account  boundary  conditions  (\ref{III:6}),  we conclude that the total
electric charge of a field configuration with a finite energy vanishes:
\begin{equation}
Q = eQ_{\phi} + qQ_{\chi} = 0.                                    \label{III:9}
\end{equation}

Substituting the power expansions for  the  functions $a_{0}\left(x\right)$, $f
\left(x\right)$, and $s\left(x\right)$ into Eqs.~(\ref{III:1}) -- (\ref{III:3}),
we  obtain  the asymptotic form of the solution as  $x \rightarrow0$:
\begin{subequations}\label{III:10}
\begin{eqnarray}
a_{0}\left( x\right)  &=&a_{0}+a_{1}x+\frac{a_{2}}{2!}x^{2}+O\left(
x^{3}\right),  \label{III:10a} \\
f_{0}\left( x\right)  &=&f_{0}+f_{1}x+\frac{f_{2}}{2!}x^{2}+O\left(
x^{3}\right),  \label{III:10b} \\
s_{0}\left( x\right)  &=&s_{0}+s_{1}x+\frac{s_{2}}{2!}x^{2}+O\left(
x^{3}\right),                                                   \label{III:10c}
\end{eqnarray}
\end{subequations}
where the next-to-leading coefficients
\begin{subequations}\label{III:11}
\begin{eqnarray}
a_{2} &=& 2 a_{0} \left(e^{2} f_{0}^{2} + q^{2}s_{0}^{2}\right)-
2 q \omega s_{0}^{2},                                          \\
f_{2} &=&f_{0}\left( m_{\phi }^{2}-g_{\phi }f_{0}^{2}+
h_{\phi}f_{0}^{4}-e^{2}a_{0}^{2}+\lambda s_{0}^{2}\right),     \\
s_{2} &=&s_{0}\left( m_{\chi }^{2}-(\omega - q a_{0})^{2}
- g_{\chi}s_{0}^{2} + h_{\chi }s_{0}^{4} \right.  \nonumber    \\
& & \left. + \lambda f_{0}^{2} \right)
\end{eqnarray}
\end{subequations}
are  determined  in  terms of  the three leading coefficients $a_{0}$, $f_{0}$,
$s_{0}$  and the model's parameters.
The next coefficients $a_{n}$,\, $f_{n}$,\, $s_{n}$, where $n = 3, 4, 5, \dots$
are  determined  by  the  six  leading  coefficients $a_{0}$, $f_{0}$, $s_{0}$,
$a_{1}$, $f_{1}$, $s_{1}$,  and the model's parameters.
It can be easily shown that if the coefficients $a_{1}$, \,$f_{1}$, and $s_{1}$
vanish, all the other coeffients  with  an  odd $n$ also vanish, and we have an
even solution of Eqs.~(\ref{III:1}) -- (\ref{III:3}).

Linearization of Eqs.~(\ref{III:1}) -- (\ref{III:3}) at large $x$ together with
corresponding boundary conditions (\ref{III:6}) lead  us to the asymptotic form
of the solution as $x \rightarrow \pm \infty$:
\begin{subequations}\label{III:12}
\begin{eqnarray}
f(x) &\sim &f_{\pm \infty} \exp \left( \mp \widetilde{m}_{\phi
\pm }x\right),                                                \\
s\left( x\right)  &\sim &s_{\pm \infty} \exp \left( \mp
\widetilde{m}_{\chi \pm }x\right),                            \\
a_{0}\left( x\right)  &\sim &a_{\pm \infty} + a_{\pm \infty}
\frac{e^{2}f^{2}_{ \pm \infty } }{2\widetilde{m}
_{\phi \pm }^{2}}          \nonumber                          \\
&&\times \exp \left( \mp 2\widetilde{m}_{\phi \pm }x\right)-
\left(\omega - q a_{\pm \infty } \right) \nonumber            \\
&&\times \frac{q s^{2}_{ \pm \infty } }{2\widetilde{m}_{\chi \pm
}^{2}}\exp \left( \mp 2\widetilde{m}_{\chi \pm }x\right),
\end{eqnarray}
\end{subequations}
where the mass  parameters $\widetilde{m}_{\phi \pm }$ and $\widetilde{m}_{\chi
\pm}$ are defined by the relations:
\begin{eqnarray}
\widetilde{m}_{\phi \pm}^{2} &=&m_{\phi}^{2} - e^{2}a_{\pm \infty}^{2},
                                                                \label{III:13a}
\\
\widetilde{m}_{\chi \pm }^{2} &=&m_{\chi }^{2} - \left(\omega -
q a_{ \pm \infty } \right)^{2}.                                 \label{III:13b}
\end{eqnarray}
From Eqs.~(\ref{III:13a}) and (\ref{III:13b}),  we  obtain the upper boundaries
on the absolute values of $a_{0}\left( \pm \infty \right) = a_{\pm \infty}$ and
$\omega$:
\begin{equation}
\left\vert a_{0}\left( \pm \infty \right) \right\vert <\frac{m_{\phi }}{e}
,\;\;\left\vert \omega\right\vert<m_{\chi }+\frac{q}{e}m_{\phi}. \label{III:14}
\end{equation}

From  Eqs.~(\ref{III:10}) -- (\ref{III:12}),  it  follows that there may be two
types  of solutions: the  symmetric one for which $f\left(-x\right) = f\left( x
\right)$,  $s\left( -x \right) = s\left( x \right)$,  $a_{0}\left( -x \right) =
a_{0}\left( x \right)$  and  the  nonsymmetric  one  that does not possess this
property.
For a symmetric solution, the series coefficients $a_{n}$, $f_{n}$, and $s_{n}$
with an odd $n$  vanish, and so in  Eqs.~(\ref{III:12}) -- (\ref{III:13b}), the
asymptotic  parameters  corresponding  to  $x\rightarrow - \infty$ are equal to
those corresponding to $x \rightarrow \infty$.

If the values of the model's  parameters  are  fixed, then  the  behavior  of a
nonsymmetric solution $f\left(x\right)$, $s\left(x\right)$, $a_{0}\left(x\right
)$ as $x \rightarrow 0$  is  determined by the six parameters $a_{0}$, $f_{0}$,
$s_{0}$, $a_{1}$, $f_{1}$, and $s_{1}$ in Eqs.~(\ref{III:10}).
The behavior of the nonsymmetric solution as $x \rightarrow \pm \infty$ is also
determined by the six parameters in Eqs.~(\ref{III:12}), namely  $a_{-\infty}$,
$f_{-\infty}$, and $s_{-\infty}$ as $x \rightarrow - \infty$  and $a_{\infty}$,
$f_{\infty}$, and $s_{ \infty}$ as $x \rightarrow \infty$.
Thus we have twelve free parameters in all.
The continuity condition for $f\left(x\right)$, $s\left(x\right)$, $a_{0}\left(
x\right)$ and their derivatives $f'\left(x\right)$, $s'\left(x\right)$, $a_{0}'
\left(x\right)$ at arbitrary $x < 0$ give us six equations.
A similar condition at arbitrary $x >0$ provides us with another six equations.
Therefore, we shall have twelve equations for determining the twelve parameters.
According to \cite{Rubakov}, this fact is an argument in favor of the existence
of the nonsymmetric solution  for  the  boundary  value  problem  in some range
of the model's parameters.
Of course, similar arguments can also be applied to a symmetric solution.

Any solution  of  field equations (\ref{II:8a}) -- (\ref{II:8c}) is an extremum
of the action $S = \int \mathcal{L} dx dt$.
At the same time, the Lagrangian density (\ref{II:1}) does  not  depend on time
in the case of field configurations (\ref{II:21}).
It  follows   that   any   solution   of   Eqs.~(\ref{III:1}) -- (\ref{III:3}),
satisfying boundary conditions (\ref{III:6}), is an extremum of  the Lagrangian
$L = \int \mathcal{L} dx$.
Let $a_{0}\left(x\right)$, $f\left(x\right)$, and $s\left(x\right)$ be a solution
of  system  (\ref{III:1})  --  (\ref{III:3}),  satisfying  boundary  conditions
(\ref{III:6}).
After  the  scale  transformation  of  the  solution's  argument $x \rightarrow
\lambda x$, the  Lagrangian  $L$  becomes  a  function  of  the scale parameter
$\lambda$.
The function $L \left( \lambda\right)$ has an extremum at $\lambda = 1$, so its
derivative with respect to $\lambda$ vanishes at this point: $\left.dL/d\lambda
\right\vert_{\lambda = 1 } = 0$.
From  this  equation, we  obtain  the  virial relation for the soliton  system:
\begin{equation}
E^{\left( E\right) }+E^{\left( P\right) }-E^{\left( G\right) } - E^{\left(
T\right) } = 0,                                                  \label{III:15}
\end{equation}
where
\begin{equation}
E^{\left( E\right) } = \int \frac{a_{0}^{\prime }{}^{2}}{2}dx    \label{III:16}
\end{equation}
is the electric field's energy,
\begin{equation}
E^{\left( G\right) }=\int \left( f^{\prime }{}^{2}+s^{\prime }{}^{2}\right)dx
                                                                 \label{III:17}
\end{equation}
is the gradient part of the soliton's energy,
\begin{equation}
E^{\left( T\right) }=\int \left( \left( \omega -qa_{0}\right)
^{2}s^{2}+e^{2}a_{0}{}^{2}f^{2}\right) dx                        \label{III:18}
\end{equation}
is the kinetic part of the soliton's energy, and
\begin{equation}
E^{\left( P\right) } = \int \left( V\left( f \right) + U\left( s \right)
+W\left( f,s\right) \right) dx                                   \label{III:19}
\end{equation}
is the potential part of the soliton's energy.

The obvious equality $E=E^{\left(E\right)}+E^{\left(T\right)}+E^{\left( G\right
)}+E^{\left(P\right)}$ and virial relation (\ref{III:15}) lead to the following
representations for the soliton system's energy:
\begin{eqnarray}
E &=&2\left(E^{\left(T\right)}+E^{\left( G\right) }\right),     \label{III:20a}
\\
E &=&2\left(E^{\left(P\right)}+E^{\left( E\right) }\right).     \label{III:20b}
\end{eqnarray}
Integrating  the  term  $a_{0}^{\prime }{}^{2}/2$ in Eq.~(\ref{III:5}) by parts
and  using  Eqs.~(\ref{III:1}), (\ref{III:4}), and (\ref{III:6}), we obtain one
more representation for the energy
\begin{equation}
E=\frac{1}{2}\omega Q_{\chi }+E^{\left( G\right) }+E^{\left( P\right)},
                                                                 \label{III:21}
\end{equation}
which, in turn, leads to the relation  between  the  Noether charge $Q_{\chi}$,
the  electric  field's  energy $E^{\left(E\right)}$,  and  the  kinetic  energy
$E^{\left(T\right)}$:
\begin{equation}
\omega Q_{\chi} = 2\left( E^{\left(E\right)} + E^{\left(T\right)} \right).
                                                                 \label{III:22}
\end{equation}

\section{\label{sec:IV} The thick-wall and thin-wall regimes of the soliton system}

In this  section, we  research  properties of the symmetric soliton solution in
two extreme regimes.
In the  thick-wall  regime,  the  mass  parameters  $\widetilde{m}_{\phi}$  and
$\widetilde{m}_{\chi}$  tend  to  zero, leading to  a  spatial spreading of the
soliton system.
This fact  and  Eqs.~(\ref{III:13a}) and (\ref{III:13b}) lead  to  the limiting
values of  the  potential $a_{0}\left(\infty\right)$  and  the  phase frequency
$\omega$ in the thick-wall regime:
\begin{equation}
\left\vert a_{0}\left( \infty \right)\right\vert = \frac{m_{\phi }}{e},\;
\omega_{\text{tk}} = \text{sgn}\left( a_{0}\left( \infty \right)
\right) \left(m_{\chi }+\frac{q}{e}m_{\phi }\right).               \label{IV:1}
\end{equation}
In the thick-wall regime, where $\widetilde{m}_{\phi }^{2} \approx \widetilde{m
}_{\chi }^{2}\rightarrow 0$, we undertake  the  following  scale transformation
of the fields and the $x$-coordinate:
\begin{eqnarray}
f\left( x\right)  &=&\Delta \bar{f}\left( \bar{x}\right)
,\qquad \quad \quad s\left( x\right) = \Delta \bar{s}\left( \bar{x}\right),
\; \nonumber \\
a_{0}\left( x\right)  &=&\frac{m_{\phi }}{e}+\frac{\Delta ^{2}}{m_{\phi }^{2}
}\bar{a}_{0}\left( \bar{x}\right),\; \quad x = \Delta^{-1}\bar{x}, \label{IV:2}
\end{eqnarray}
where the scale factor $\Delta$ is defined as
\begin{eqnarray}
\Delta^{2} & = &m_{\phi}^{2} - e^{2} a_{0}^{2}\left(\infty\right)
\approx m_{\chi }^{2} - \left(\omega - q a_{0}\left(\infty\right)
\right)^{2} \nonumber \\
&\approx &\kappa ^{2}\left( \omega _{\text{tk}}^{2}-\omega ^{2}\right).
                                                                   \label{IV:3}
\end{eqnarray}
In Eq.~(\ref{IV:3}),  the  factor $\kappa$  is expressed in terms of the scalar
particles' masses and the gauge coupling constants:
\begin{equation}
\kappa =e\left[ \frac{m_{\phi }m_{\chi }}{\left( e m_{\phi }+q m_{\chi
}\right) \left(e m_{\chi}+q m_{\phi}\right)}\right]^{\frac{1}{2}}. \label{IV:4}
\end{equation}

Let us consider the functional $F$, which has been defined in Eq.~(\ref{II:12}).
This functional  is  related  to  the  energy  functional  by means of Legendre
transformation: $F\left(\omega\right)=E\left(Q_{\chi}\right)-\omega Q_{\chi }$.
On  field  configuration  (\ref{IV:2}), the  functional $F$  can  be written as
\begin{equation}
F\left(\omega\right)=\Delta^{3}\bar{F} + O\left(\Delta^{5}\right), \label{IV:5}
\end{equation}
where the functional $\bar{F}$ does not depend on $\omega$:
\begin{eqnarray}
\bar{F} &=&\int \left[ \bar{f}^{\prime }\left( \bar{x}%
\right) ^{2}+\bar{s}^{\prime}\left( \bar{x}\right) ^{2}+\bar{%
f}\left( \bar{x}\right) ^{2}+\bar{s}\left( \bar{x}\right)
^{2}\right. \label{IV:6} \\
&&\left. -\frac{g_{\phi }}{2}\bar{f}\left( \bar{x}\right)^{4}-%
\frac{g_{\chi }}{2}\bar{s}\left( \bar{x}\right) ^{4}+\lambda
\bar{f}\left( \bar{x}\right) ^{2}\bar{s}\left( \bar{x}%
\right) ^{2}\right] d\bar{x}. \nonumber
\end{eqnarray}
In the thick-wall regime, the phase frequency  $\omega$  tends to  the limiting
value  $\omega_{\text{tk}}$,  so  the  parameter  $\Delta$  vanishes, and it is
possible  to  ignore  the  higher-order  terms in $\Delta$ in Eq.~(\ref{IV:5}).
Using  known  properties  of  Legendre  transformation,  we obtain sequentially
\begin{eqnarray}
Q_{\chi }\left( \omega \right)& = & -\frac{dF\left( \omega \right) }
{d\omega } = 3\bar{F}\kappa^{3}\omega \left(\omega_{\text{tk}}^{2} - \omega
^{2}\right) ^{\frac{1}{2}},                                        \label{IV:7}
\\
E\left( \omega \right)  &=&F\left( \omega \right) - \omega \frac{dF\left(
\omega \right) }{d\omega } \nonumber \\
&=&\bar{F}\kappa ^{3}\left( 2\omega ^{2}+\omega _{\text{tk}
}^{2}\right) \left( \omega _{\text{tk}}^{2}-
\omega^{2}\right)^{\frac{1}{2}}.                                   \label{IV:8}
\end{eqnarray}
From  Eqs.~(\ref{IV:7})  and  (\ref{IV:8}),  we  obtain  the  dependence of the
energy $E$ on the Noether charge $Q_{\chi}$ in the thick-wall regime:
\begin{equation}
E\left(Q_{\chi }\right)=\omega_{\text{tk}}Q_{\chi }-\frac{1}{54}\frac{1}
{\bar{F}^{2}\kappa ^{6}\omega_{\text{tk}}^{3}}Q_{\chi }^{3}
+O\left(Q_{\chi}^{5}\right).                                       \label{IV:9}
\end{equation}

We see from Eqs.~(\ref{III:9}),  (\ref{IV:7}),  (\ref{IV:8}),  and (\ref{IV:9})
that the energy $E$ and  the  Noether  charges $Q_{\phi}$ and $Q_{\chi}$ of the
soliton system tend to zero in the thick-wall regime.
Further,  Eq.~(\ref{IV:9}),  basic  relation (\ref{II:22}), and  the inequality
$\omega^{2} < \omega_{\text{tk}}^{2}$  lead to the conclusion that $E\left( Q_{
\chi }\right) < \omega_{\text{tk}}Q_{\chi }$  for  all  values  of  $Q_{\chi}$.
From  Eqs.~(\ref{III:8b}), (\ref{III:9}),  and  (\ref{IV:1})  it  follows  that
$\omega_{\text{tk}}Q_{\chi}$ is equal to $m_{\phi}\left\vert Q_{\phi}\right\vert
+m_{\chi}\left\vert Q_{\chi}\right\vert$, which, in turn, is the rest energy of
the neutral  plan-wave  configuration formed from the charged scalar $\phi$ and
$\chi$-particles.
Hence, the symmetric  soliton  system is stable to decay into the scalar $\phi$
and $\chi$-particles.

The  second  extremal regime  of  the symmetric soliton system is the thin-wall
regime in which the absolute value of the phase frequency tends to some minimum
value $\omega_{\text{tn}}$.
In the  thin-wall  regime,  the  spatial  size  of the soliton system increases
indefinitely, with the result that its energy $E$ and Noether charges $Q_{\phi}$
and $Q_{\chi}$ also tend to infinity.
In  the  thin-wall  regime,  when  the  spatial  size  of  the  soliton  system
$L \rightarrow \infty$,  the  gradient  operator gives a factor proportional to
$L^{-1}$.
Therefore,  we  can  ignore  the electric field's energy (\ref{III:16}) and the
gradient   energy   (\ref{III:17})   in   comparison  with  the  kinetic energy
(\ref{III:18}) and the potential energy (\ref{III:19}).
Then, from  Eq.~(\ref{III:15}) it follows that  the following limiting relation
holds in the thin-wall regime:
\begin{equation}
\underset{\omega \rightarrow \omega_{\text{tn}}}{\lim}
\frac{E^{\left(T\right) }}{E^{\left( P\right) }} = 1,             \label{IV:10}
\end{equation}
and, as a consequence,
\begin{equation}
\underset{\omega \rightarrow \omega_{\text{tn}}}{\lim}
\frac{2E^{\left(T\right) }}{E} =
\underset{\omega \rightarrow \omega_{\text{tn}}}{\lim }
\frac{2E^{\left( P\right) }}{E} = 1.                              \label{IV:11}
\end{equation}
Further,  electric  charge density (\ref{III:4}) tends to zero in the thin-wall
regime, since the soliton system's  electric  charge is strictly equal to zero,
whereas its spatial size tends to infinity.
Then, using  Eqs.~(\ref{III:4}), (\ref{III:18}),  and  (\ref{IV:11}), we obtain
the limiting relation
\begin{equation}
\underset{\omega \rightarrow \omega_{\text{tn}}}{\lim }
\frac{2E^{\left(T\right)}}{Q_{\chi }} =
\underset{\omega \rightarrow \omega_{\text{tn}}}{\lim }
\frac{E}{Q_{\chi }} =
\omega _{\text{tn}},                                              \label{IV:12}
\end{equation}
which is  consistent  with basic relation (\ref{II:22}) and Eq.~(\ref{III:22}).

\section{\label{sec:V} Numerical results}

The  system  of  differential  equations  (\ref{III:1})  --  (\ref{III:3}) with
boundary conditions (\ref{III:6})  is  the  mixed boundary value problem on the
infinite  interval  $x  \in  \left( -\infty, \infty \right)$.
This  boundary  value  problem can   be   solved  only  by  numerical  methods.
In  this  paper, the boundary value problem  was  solved using the {\sc{Maple}}
package \cite{maple}  by  the  method  of  finite  differences  and  subsequent
Newtonian iterations.
Equations (\ref{II:22}), (\ref{III:9}), and  (\ref{III:15})  were used to check
the correctness of numerical solutions.

Let us  discuss  possible  types  of  solutions  of the boundary value problem.
If the  quartic  coupling  constant $\lambda$  and the electromagnetic coupling
constants $e$ and $q$  are set  equal to zero, then the Lagrangian (\ref{II:1})
will describe the  system  of  two self-interacting complex scalar fields that,
however, do not interact with each other.
In this case, the  boundary  value problem has the solution describing a system
of two non-interacting non-gauged one-dimensional Q-balls.
Generally, these  two Q-balls  have different shapes and can be at an arbitrary
distance  from  each other, so the solution will not be symmetric. 
However, the situation changes when  the  electromagnetic interaction is turned
on.
In this case, from Eq.~(\ref{III:9}) it  follows  that  the electric charges of
two Q-ball components are equal in magnitude, but opposite in sign.
It is important to  note  that  the  electric charges of two gauged Q-balls are
conserved separately owing to the neutrality of the Abelian gauge field.
Since  the  opposite  electric  charges  attract  each  other,  the   initially
nonsymmetric soliton system transits to a symmetric one.
Now we turn on the quartic  interaction  between  the two complex scalar fields
$\phi$ and $\chi$ by  letting  the coupling constant $\lambda$ be some positive
value.
From Eq.~(\ref{II:4}),  it  follows that the energy of the quartic  interaction
increases  with  the increase of overlap between  the  Q-ball components of the
soliton system and  is  negligible  at  large  separations  between  the Q-ball
components.
Such a  behavior  of  the  quartic  interaction  corresponds  to  a short-range
repulsive force  between  the  Q-ball  components,  while  the  electromagnetic
long-range  attractive  force  results  in  the  confinement of the the  Q-ball
components.
One would  expect that for a  sufficiently  large  positive  coupling  constant
$\lambda$,  the  action  of  these  opposite  forces  leads  to  an equilibrium
nonsymmetric  soliton   configuration,   which   is  the  solution of  boundary
value problem (\ref{III:1})  --  (\ref{III:3}), and (\ref{III:6}).
Indeed,  we  shall  see  later that such a nonsymmetric soliton solution really
exists.

The system of differential  equations (\ref{III:1}) -- (\ref{III:3}) depends on
the  ten  dimensional  parameters: $\omega$, $e$,  $q$, $m_{\phi}$, $m_{\chi}$,
$g_{\phi}$, $g_{\chi}$, $h_{\phi}$, $h_{\chi}$, and $\lambda$.
It is readily  seen,  however, that  the  dimensionless  functions $a_{0}\left(
x\right)$, $f\left(x\right)$,  and  $s\left(x\right)$  can  depend only on nine
independent dimensionless combinations of these parameters.
Therefore without loss of generality, we  can choose the mass $m_{\phi}$ of the
scalar $\phi$-particle as the energy unit.
We consider  a general case in which the corresponding dimensionless parameters
are values of the same order: $\tilde{e}=e/m_{\phi}=0.2$, $\tilde{q}=q/m_{\phi}
=0.2$, $\tilde{m}_{\chi}=m_{\chi}/m_{\phi}=1.25$, $\tilde{g}_{\phi}=g_{\phi}/m_{
\phi}^{2}=1$,  $\tilde{g}_{\chi}=g_{\chi}/m_{\phi}^{2}=1.5$, $\tilde{h}_{\phi}=
h_{\phi}/m_{\phi}^{2} = 0.22$,  $\tilde{h}_{\chi }  =  h_{\chi }/m_{\phi}^{2} =
0.31$, and $\tilde{\lambda} = \lambda /m_{\phi }^{2} = 0.2$.

\begin{figure}
\includegraphics{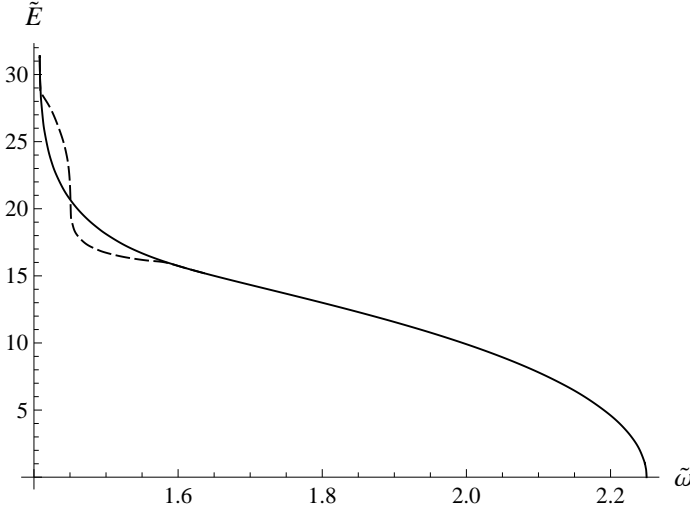}
\caption{\label{fig1}  The  dependence  of  the  dimensionless  soliton  energy
$\tilde{E}   =   m_{\phi}^{-1} E$   on   the  dimensionless   phase   frequency
$\tilde{\omega} = m_{\phi}^{-1} \omega$.  The  solid  curve  corresponds to the
symmetric soliton  system, and the dashed curve corresponds to the nonsymmetric
one.}
\end{figure}
Figures~1 and 2 present the dependences of  the soliton's dimensionless  energy
$\tilde{E}=m_{\phi}^{-1} E$ and Noether charge $Q_{\chi}$ on the  dimensionless
phase  frequency  $\tilde{\omega} =  m_{\phi}^{-1} \omega$.
The most striking feature of Figs.~1 and 2 is  the coexistence of the symmetric
and nonsymmetric soliton solutions.
Indeed, it has been found numerically that the symmetric soliton solution exists
in the range from  the minimum value $\tilde{\omega}_{\min} = 1.4079916$, which
we managed to  reach  by numerical methods, to the maximum value $\tilde{\omega
}_{\text{tk}} = 2.25$.
On the contrary, the nonsymmetric soliton solution  exists only in the interval
from  the  left  bifurcation  piont $\tilde{\omega}_{\text{lb}} = 1.409$ to the
right one $\tilde{\omega}_{\text{rb}} = 1.601$.
We also  see  that  the   two   types   of   curves   have  intersection points
at $\tilde{\omega}_{\text{i1}}$ and $\tilde{\omega}_{\text{i2}}$ in Figs.~1 and
2, respectively.
These intersection points are slightly different: $\tilde{\omega}_{\text{i1}} =
1.4504025$, whereas $\tilde{\omega}_{\text{i2}} = 1.4504280$.
In each of the figures, the  solid  and  dashed  curves  bound the two regions,
which connect at the intersection points.
Using Eq.~(\ref{II:22}),  it  can  easily  be  shown  that  the  areas of these
regions are equal to each other, so we have the relations:
\begin{equation}
\int\nolimits_{\tilde{\omega}_{\text{lb}}}^{\tilde{\omega}_{\text{rb}}}
\left[ Q_{\chi \text{a}}\left(\tilde{\omega} \right) -
Q_{\chi \text{s}}\left( \tilde{\omega} \right) \right]
d\tilde{\omega} = 0,                                                \label{V:1}
\end{equation}
and
\begin{equation}
\int\nolimits_{\tilde{\omega}_{\text{lb}}}^{\tilde{\omega}_{\text{rb}}}
\left[ \tilde{E}_{\text{a}}\left(\tilde{\omega} \right) -
\tilde{E}_{\text{s}}\left( \tilde{\omega} \right) \right]
d\tilde{\omega} = 0,                                                \label{V:2}
\end{equation}
which were  checked numerically.
\begin{figure}
\includegraphics{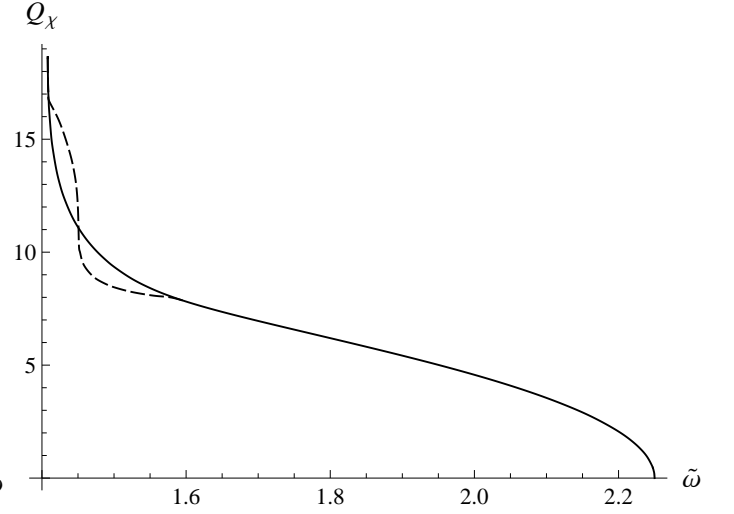}
\caption{\label{fig2}  The dependence of the  soliton Noether charge $Q_{\chi}$
on  the  dimensionless phase frequency $\tilde{\omega} = m_{\phi}^{-1} \omega$.
The solid  curve corresponds to the  symmetric soliton system,  and  the dashed
curve corresponds to the nonsymmetric one.}
\end{figure}

When $\tilde{\omega}$ tends to its  minimal value $\tilde{\omega}_{\text{tn}}$,
the symmetric soliton system goes into the thin-wall regime.
In this regime, the energy $\tilde{E}$, the Noether charges $Q_{\chi}$ and $Q_{
\phi}$, and  the  effective  spatial  size  $L$ of the symmetric soliton system
increase indefinitely.
In particular, we found  numerically  that $\tilde{E}$, $Q_{\chi}$, $Q_{\phi}$,
and $L$ increase logarithmically as $\tilde{\omega}  \rightarrow  \tilde{\omega
}_{\text{tn}}$:
\begin{eqnarray}
\tilde{E} &\sim & - \tilde{\omega}_{\text{tn}} B \ln \left( \tilde{\omega} -
\tilde{\omega}_{\text{tn}}\right), \label{V:3a}
\\
Q_{\chi } &\sim & - B\ln\left(\tilde{\omega}-\tilde{\omega}_{\text{tn}}\right),
\label{V:3b}
\\
Q_{\phi } &\sim &   B \frac{q}{e}\ln\left( \tilde{\omega} - \tilde{\omega}_{
\text{tn}}\right), \label{V:3c}
\\
L &\sim & - C \ln\left( \tilde{\omega} - \tilde{\omega}_{\text{tn}}\right),
\label{V:3d}
\end{eqnarray}
where $B$ and $C$ are some positive constants, and the limiting thin-wall phase
frequency $\tilde{\omega}_{\text{tn}} = 1.4079869$.
Note that this numerical estimation of $\tilde{\omega}_{\text{tn}}$ is slightly
less  than  the  minimal  value  $\tilde{\omega}_{\min} = 1.4079916$, which was
reached by numerical methods.
Note also that in the thin-wall regime,  the  behavior of $E$, $Q_{\chi}$, $Q_{
\phi}$, and $L$   is  similar  to  that  of  the  corresponding  values  of the
one-dimensional non-gauged Q-ball,  as  it  follows  from  Eqs.~(\ref{V:3a}) --
(\ref{V:3d}) and (\ref{A1:VIII}) -- (\ref{A1:IXa}).

When $\tilde{\omega}$ tends to its  maximal value $\tilde{\omega}_{\text{tk}}$,
the symmetric soliton system goes into the thick-wall regime.
In this regime, the  soliton  system  is spread out over one-dimensional space,
while the amplitudes of the scalar fields $\phi$  and  $\chi$  tend  to zero as
$\left(\tilde{\omega}_{\text{tk}} - \tilde{\omega}\right)^{1/2}$  in accordance
with Sec.~\ref{sec:IV}.
It was found numerically that in the thick-wall regime, $\tilde{E}$, $Q_{\chi}$,
and $Q_{\phi}$ also tend to zero as $\left(\tilde{\omega}_{\text{tk}} - \tilde{
\omega}\right)^{1/2}$, whereas  the  effective  spatial  size  $L$  diverges as
$\left(\tilde{\omega}_{\text{tk}} - \tilde{\omega}\right)^{-1/2}$:
\begin{eqnarray}
\tilde{E} &\sim &b\,\tilde{\omega}_{\text{tk}}\left(\tilde{\omega}_{\text{tk}}-
\tilde{\omega}\right)^{\frac{1}{2}},                               \label{V:4a}
\\
Q_{\chi } &\sim &b\left( \tilde{\omega}_{\text{tk}} - \tilde{\omega}
\right)^{\frac{1}{2}},                                             \label{V:4b}
\\
Q_{\phi } &\sim &-b \frac{q}{e}\left( \tilde{\omega}_{\text{tk}} -
\tilde{\omega}\right)^{\frac{1}{2}},                               \label{V:4c}
\\
L &\sim & c \left(\tilde{\omega}_{\text{tk}} - \tilde{\omega}\right)^{-1/2}.
                                                                   \label{V:4d}
\end{eqnarray}
From Eqs.~(\ref{V:4a}) -- (\ref{V:4d})  and  (\ref{A1:V}) -- (\ref{A1:VIa}), it
follows that the behavior of $E$, $Q_{\chi}$, $Q_{\phi}$, and  $L$  is  similar
to that of the corresponding values of the one-dimensional non-gauged Q-ball in
the thick-wall regime.

\begin{figure}
\includegraphics{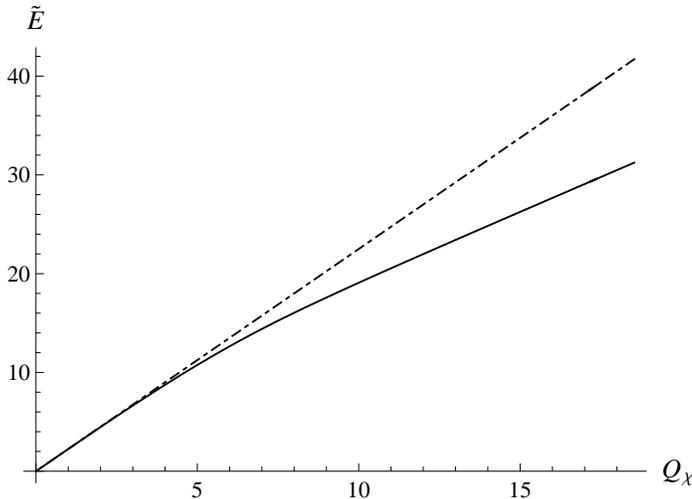}
\caption{\label{fig3}  The dependence  of  the  dimensionless energy $\tilde{E}
=  m_{\phi}^{-1} E$  of  the  symmetric  soliton  system on  the Noether charge
$Q_{\chi}$  (solid curve). The dash-dotted line is the straight line $\tilde{E}
= \tilde{\omega}_{\text{tk}} Q_{\chi}  =  \left(1  +  m_{\chi}/m_{\phi} \right)
Q_{\chi}$.}
\end{figure}
Figure~\ref{fig3} shows the dependence of  the dimensionless energy $\tilde{E}$
of  the   symmetric   soliton   system   on   the  Noether  charge  $Q_{\chi}$.
We see  that  the  dependence $\tilde{E}\left(Q_{\chi}\right)$ is an increasing
convex $(d\tilde{E}/dQ_{\chi} > 0,\,d^{2}\tilde{E}/dQ_{\chi}^{2} < 0)$ function
outgoing from the coordinate origin.
It follows that  the  symmetric  soliton system is stable to decay into several
smaller ones.
We also see that in accordance  with  Sec.~\ref{sec:IV}, the  curve  $\tilde{E}
\left(Q_{\chi}\right)$ lies below the straight  line $\tilde{E} = \tilde{\omega
}_{\text{tk}}Q_{\chi}$ for all positive $Q_{\chi}$.
From this, it follows that the symmetric soliton system is stable to decay into
massive scalar $\phi$ and $\chi$-bosons.
\begin{figure}
\includegraphics{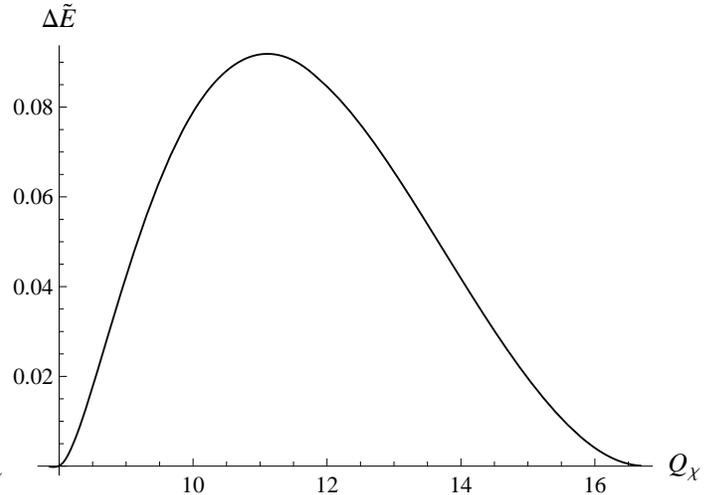}
\caption{\label{fig4}   The dependence of the energy difference $\Delta \tilde{
E} = \tilde{E}_{\text{s}} - \tilde{E}_{\text{a}}$  between  the  symmetric  and
nonsymmetric  soliton  solutions  on the Noether charge $Q_{\chi}$.}
\end{figure}

\begin{figure}[b]
\includegraphics{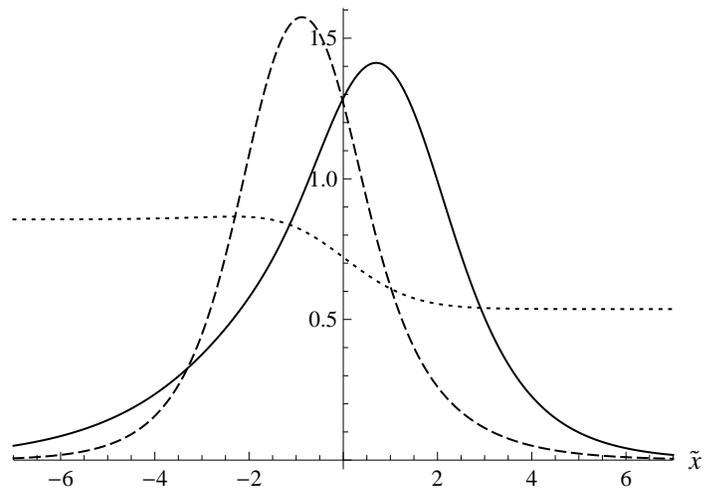}
\caption{\label{fig5} The nonsymmetric numerical solution for $f\left(\tilde{x}
\right)$ (solid curve), $s\left(\tilde{x}\right)$  (dashed curve), and $\tilde{
e} a_{0}\left( \tilde{x} \right)$  (dotted curve).    The  dimensionless  phase
frequency $\tilde{\omega} = 1.5 $. }
\end{figure}
In Fig.~\ref{fig4}, we  can see the dependence of the energy difference $\Delta
\tilde{E}=\tilde{E}_{\text{s}} -\tilde{E}_{\text{a}}$ between the symmetric and
nonsymmetric  soliton  solutions  on the Noether charge $Q_{\chi}$.
From Fig.~\ref{fig4}, it follows that the energy of the symmetric soliton system
slightly exceeds the energy of the nonsymmetric one in the whole  range  of the
Noether  charge  $Q_{\chi}$   for  which  the  existence  of  the  nonsymmetric
soliton system is possible.
It follows that the nonsymmetric soliton system  is  more  preferable  from the
viewpoint of energy as compared to the symmetric one.
Note, however, that the difference $\Delta \tilde{E}$ is rather small and is of
the order of $0.1 \%$  of the soliton system's energy.
As well  as  the  symmetric soliton  system,  the nonsymmetric one is stable to
decay into massive scalar bosons.
Note, however, that the symmetric soliton system is unstable to the  transition
into the nonsymmetric one  because  of  the  possibility of tunneling under the
potential barrier.
\begin{figure}
\includegraphics{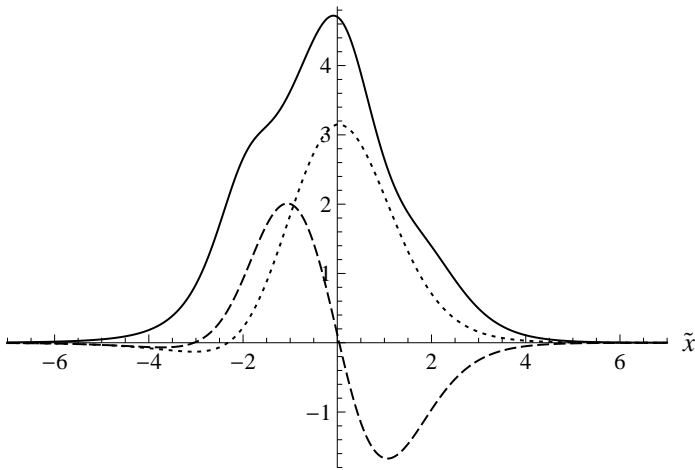}
\caption{\label{fig6} The dimensionless versions of the energy density $\tilde{
\mathcal{E}}  =  m_{\phi}^{-2} \mathcal{E}$  (solid curve), the scaled electric
charge  density $\tilde{e}^{-1}\tilde{j_{0}}=\tilde{e}^{-1} m_{\phi}^{-2}j_{0}$
(dashed curve), and the  scaled electric field strength $\tilde{e}^{-1}\tilde{E
}_{x} = \tilde{e}^{-1} m_{\phi}^{-1} E_{x}$  (dotted  curve), corresponding  to
the nonsymmetric solution in Fig.~5.}
\end{figure}

Figure~\ref{fig5} presents the  nonsymmetric  soliton solution corresponding to
the dimensionless phase frequency $\tilde{\omega}=1.5$, whereas Fig.~\ref{fig6}
presents the energy  and  electric  charge  densities  and  the  electric field
strength corresponding to Fig.~\ref{fig5}.
The  nonsymmetric  character   of   the   soliton   system   is   obvious  from
Figs.~\ref{fig5} and  \ref{fig6}.
The most interesting feature of the nonsymmetric soliton system is the presence
of the unidirectional electric field in its interior, as for a plane capacitor.
From  Fig.~\ref{fig5}, it   follows   that   the   charged  scalar  $\phi$  and
$\chi$-particles can acquire the energy equal to $ -e \Delta a_{0} \approx 0.32
m_{\phi}$   in   the   electric  field  of  the  nonsymmetric  soliton  system.
Note that  this  energy  is  comparable  with  the  scalar  particles'  masses.
Lighter    particles  (e.g.  light  charged   fermions)  passing   through  the
interior  of   the   nonsymmetric   soliton   system   can  be  accelerated  to
relativistic velocities and energies.

\begin{figure}[b]
\includegraphics{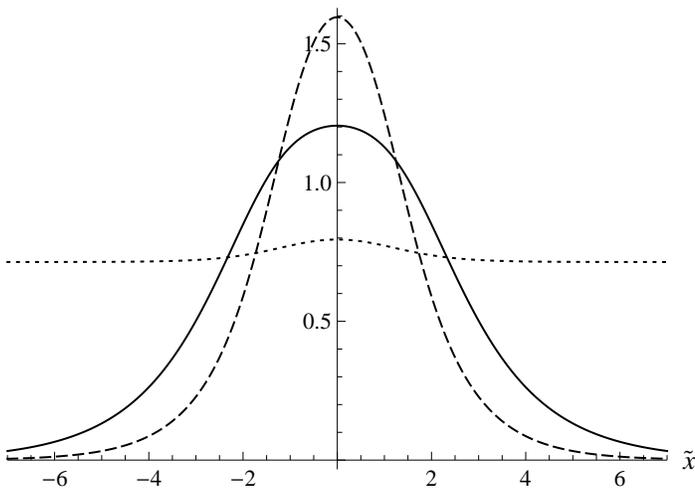}
\caption{\label{fig7}  The  symmetric  numerical solution for $f\left(\tilde{x}
\right)$ (solid curve), $s\left(\tilde{x}\right)$ (dashed curve), and $\tilde{e}
a_{0}\left(\tilde{x}\right)$ (dotted curve).  The dimensionless phase frequency
$\tilde{\omega} = 1.5 $. }
\end{figure}
\begin{figure}
\includegraphics{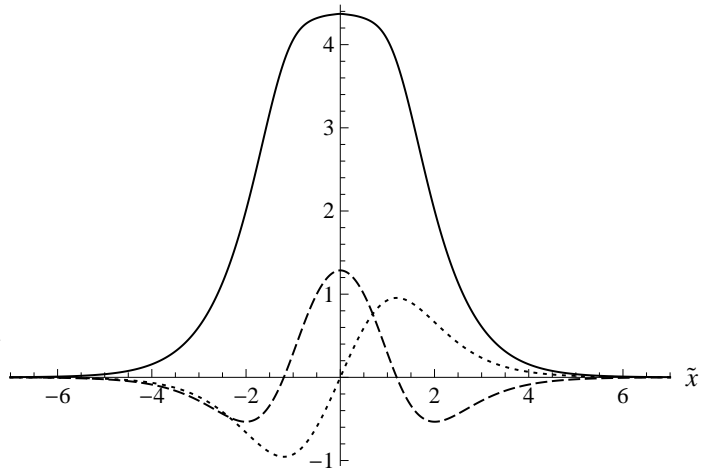}
\caption{\label{fig8} The dimensionless versions of the energy density $\tilde{
\mathcal{E}}  =  m_{\phi}^{-2} \mathcal{E}$  (solid curve), the scaled electric
charge  density $\tilde{e}^{-1}\tilde{j_{0}}=\tilde{e}^{-1} m_{\phi}^{-2}j_{0}$
(dashed curve), and the  scaled electric field strength $\tilde{e}^{-1}\tilde{E
}_{x} = \tilde{e}^{-1}m_{\phi}^{-1} E_{x}$ (dotted curve), corresponding to the
symmetric solution in Fig.~7.}
\end{figure}
Similar to Figs.~\ref{fig5} and \ref{fig6}, Figs.~\ref{fig7} and \ref{fig8} give
information about the symmetric soliton solution.
From Fig.~\ref{fig8}, it follows that the energy and electric  charge densities
are  symmetric  with  respect  to  the  center of the soliton system, while the
electric field strength is  antisymmetric.
For positive $\omega$, it is directed from  the soliton  system's center, so it
attracts negatively  charged  particles  and  repels  positively  charged ones.
For negative $\omega$,  it  is  directed   to  the soliton  system's center, so
the  roles  of  negatively  and  positively charged particles are interchanged.
For positive (negative) $\omega$, the form  of  the  electromagnetic  potential
$a_{0}$ corresponds  to  a  potential  well for negatively (positively) charged
particles.
It follows  that  bound  fermionic and bosonic states can exist in the electric
field of the symmetric soliton system.

\section{\label{sec:VI} Conclusion}

In  the  present  paper,  the  one-dimensional  nontopological  soliton  system
consisting of two self-interacting complex scalar fields has been investigated.
The scalar fields  interact with each other through the Abelian gauge field and
the quartic scalar interaction.
The finiteness  of  the  energy of the  one-dimensional soliton system leads to
its electric  neutrality,  so  its two scalar components have opposite electric
charges.
The  neutrality  of  the Abelian gauge field leads to the separate conservation
of the electric charges of these scalar components.
The interplay  between  the  attractive  electromagnetic  interaction  and  the
repulsive   quartic  interaction  leads  to  the  existence  of  symmetric  and
nonsymmetric soliton systems.

The symmetric  soliton  system  exists  in  the whole allowable interval of the
phase frequency $\omega$.
When $\omega $  tends  to  its  minimal  (maximal) value, the symmetric soliton
system goes into the thin-wall (thick-wall) regime.
In the thin-wall regime, the energy, the  Noether charges, and the spatial size
of the symmetric soliton system tend to infinity.
In the thick-wall regime, the spatial size of the symmetric soliton system also
tends to infinity, but the energy and the Noether charges tend to zero.
In contrast to  this, the  nonsymmetric  soliton  system  exists  only  in some
interior subinterval between the minimal and maximal allowable phase frequencies
$\omega_{\text{tn}}$ and $\omega_{\text{tk}}$.
It follows that there exists an interval of the Noether charge $Q_{\chi}$ (and,
consequently, an interval of the Noether charge $Q_{\phi}=-q e^{-1} Q_{\chi}$),
where the symmetric and nonsymmetric soliton systems coexist.
In all this interval, the  energy  of the nonsymmetric soliton system turns out
to be less than that of the symmetric soliton system, so the symmetric  soliton
system can turn into the nonsymmetric one through quantum tunneling.
Both symmetric and nonsymmetric soliton systems are stable to decay into massive
scalar $\phi$ and $\chi$-bosons.

Despite the fact that the soliton system is electrically neutral, it nevertheless
possesses a nonzero electric field in its interior.
Note that the electric fields of the symmetric and nonsymmetric soliton systems
are essentially different.
The electric field of the nonsymmetric soliton  system is unidirectional in its
interior, like the electric field of a plane capacitor.
It can accelerate  light  particles up to relativistic velocities and energies.
In contrast, the  electric field of the symmetric soliton system corresponds to
the electromagnetic potential of a potential well.
In such an electric field, the existence of bound bosonic and  fermionic states
is possible.

It is known \cite{lee, fried} that  the field configuration of a nontopological
soliton composed  only  of  scalar  fields  can  be  described  in  terms  of a
mechanical analogy.
For the one-dimensional case, it corresponds to the motion of  a  particle with
the unit mass in the “time” $x$   in  the conservative force field of a certain
potential.
The dimension of “space” in  which the particle moves is equal to the number of
scalar fields constituting the nontopological soliton.
Using this  analogy, one  can  easily  explain the  behavior of the pure scalar
nontopological  soliton  both  in  the thin-wall and in the thick-wall regimes.
Moreover, one  can  easily  determine whether a soliton solution can exists for
any values of the model's parameters.
At the same time, system of differential equations (\ref{III:1}) -- (\ref{III:3})
describing  the  soliton  system  of  the  present  paper has no interpretation
in terms of any mechanical analogy.
For this reason, the existence of the soliton  system should be established for
any given set of the model's parameters by means of numerical methods.

Finally,  let us  stress  the  specific  character of the $(1 + 1)$-dimensional
electromagnetic field.
Its  characteristic  feature  is  the  absence  of  nondiagonal  terms  of  the
electromagnetic stress-energy tensor.
This is because  the  magnetic  field  does not exist in  $(1 + 1)$-dimensions,
so the Poynting vector vanishes there. 
Therefore, the $(1 + 1)$-dimensional  electromagnetic  field  can  not transfer
any energy or momentum.
Instead, the scalar fields' kinetic energy can transform to the one-dimensional
electric  field's  energy,  which,  in  turn, can  transform back to the scalar
fields' energy.
Note also that in  $(1 + 1)$-dimensions, the potential energy of two oppositely
charged  particles  is  proportional  to  the  distance  between  them,  so the
electromagnetic interaction is confining there.
Thus, we  can conclude that the $(1+1)$-dimensional electromagnetic interaction
is similar to an elastic string.
The only  difference  is  that  there is no energy and momentum transfer in the
one-dimensional  electric  field,  whereas  in  the  elastic  string  waves can
transfer energy and momentum.
The  behaviour of the $(1 + 1)$-dimensional electromagnetic field is completely
determined  by  Gauss's  law,  which is not a dynamic field equation but is the
condition imposed on an initial field configuration.
Indeed, in  the  adopted  gauge $A_{x} = 0$,  Gauss's law does not contain time
derivatives of the electromagnetic potential $A_{0}$.
In this connection, it can be said that the $(1+1)$-dimensional electromagnetic
field is not a dynamic one.

\begin{acknowledgments}

The research  is  carried  out  at  Tomsk  Polytechnic  University  within  the
framework of  Tomsk  Polytechnic University Competitiveness Enhancement Program
grant.

\end{acknowledgments}

\appendix

\section{The  one-dimensional non-gauged Q-ball}

Here we  collect  formulae  concerning   the  one-dimensional non-gauged Q-ball
in  the  model of a self-interacting complex scalar  field  with  the six-order
self-interaction potential $V \left(\left\vert \phi \right\vert \right) = m^{2}
\left\vert  \phi \right\vert^{2} - g  \left\vert  \phi  \right\vert^{4}/2  +  h
\left\vert  \phi \right\vert^{6}/3$.
Note  that  an  analytical  Q-ball  solution  exists  only  in  the  $\left(1+1
\right)$-dimensional case \cite{lee}, where it can  be  written  as
\begin{eqnarray}
\phi\left(t, x\right) &=&\frac{2}{\sqrt{g}}\sqrt{m^{2}-\omega^{2}} \nonumber \\
&&\times \left( 1+\left( 1-\frac{m^{2}-\omega ^{2}}{m^{2}-
\omega _{\text{tn}}^{2}}\right) ^{\frac{1}{2}}\right.              \nonumber \\
&&\times \Biggl. \cosh \left( 2\sqrt{m^{2}-\omega ^{2}}\left( x-x_{0}\right)
\right) \Biggr) ^{-\frac{1}{2}}                                 \nonumber \\
&&\times \exp\left(-i\omega \left(t - t_{0}\right) \right).        \label{A1:I}
\end{eqnarray}
In Eq.~(\ref{A1:I}), the squared  phase frequency $\omega^{2}\in\left( \omega_{
\text{tn}}^{2},\; m^{2} \right)$, where
\begin{equation}
\omega_{\text{tn} }^{2} = m^{2}\left(1-\frac{3}{16}\frac{g^{2}}{m^2 h}\right).
                                                                  \label{A1:II}
\end{equation}
The Noether charge and the energy of the one-dimensional Q-ball can be expressed
in a rather compact form:
\begin{eqnarray}
Q &=&4\omega \sqrt{\frac{3}{h}}\text{arctanh}\left( \left[ \frac{%
m^{2}-\omega _{\text{tn} }^{2}}{m^{2}-\omega ^{2}}\right]^{\frac{1}{2}}\right.
\nonumber \\
&&\left. -\left[ \frac{m^{2}-\omega _{\text{tn} }^{2}}
{m^{2}-\omega ^{2}}-1\right]^{\frac{1}{2}}\right),               \label{A1:III}
\end{eqnarray}
and
\begin{eqnarray}
E &=&\omega Q-2\sqrt{\frac{3}{h}}\left( \omega ^{2}-\omega _{\text{tn}
}^{2}\right) \nonumber \\
&&\times \text{arctanh}\left(\left[\frac{\sqrt{m^{2}-\omega _{\text{tn} }^{2}}-
\sqrt{\omega^{2}-\omega_{\text{tn}}^{2}}}{\sqrt{m^{2}-\omega _{\text{tn}}^{2}}+
\sqrt{\omega ^{2}-\omega _{\text{tn} }^{2}}}\right] ^{\frac{1}{2}}\right)
\nonumber \\
&&+\sqrt{\frac{3}{h}}\sqrt{\left( m^{2}-\omega ^{2}\right) \left(
m^{2}-\omega _{\text{tn} }^{2}\right)}\,.                         \label{A1:IV}
\end{eqnarray}

Let us present  the expressions of the Noether charge $Q$ and the energy $E$ in
two extreme regimes.
In the thick-wall regime, the squared phase frequency tends to its maxmum value:
$\omega^{2} \rightarrow m^{2}$.
Using  Eqs.~(\ref{A1:III})  and  (\ref{A1:IV}),  we  obtain  the expressions of
the  soliton's Noether  charge  and  energy in the thick-wall regime:
\begin{eqnarray}
Q &=&\text{sgn}\left( \omega \right)
\frac{2\sqrt{6}m^{3/2}}{\sqrt{h(m^{2}-\omega_{\text{tn}}^{2})}}
\sqrt{\delta}                                                      \nonumber \\
&&\times \left(1-\frac{7 m^{2}-15 \omega_{\text{tn}}^{2}}{12 m  \left(
m^{2}- \omega_{\text{tn}}^{2}\right) }\delta + O \left(\delta
^{2}\right) \right),                                               \label{A1:V}
\end{eqnarray}
and
\begin{eqnarray}
E &=&\frac{2\sqrt{6}m^{5/2}}{\sqrt{h (m^{2}-\text{$\omega $}_{\text{tn}}^{2})}}
\sqrt{\delta }                                                    \label{A1:VI}
\\
&&\times \left( 1 - \frac{11m^{2} - 19\text{$\omega $}_{\text{tn}}^{2}}
{12 m\left(m^{2}  - \text{$\omega $}_{\text{tn}}^{2}\right) }\delta +
O \left( \delta^{2}\right) \right),  \nonumber
\end{eqnarray}
where the variable $\delta = m - \left\vert \omega \right \vert$.
Furthermore,  Eq.~(\ref{A1:I}) leads  to the soliton's width at  half-height in
the thick-wall regime:
\begin{eqnarray}
L &=&\frac{\cosh ^{-1}\left( 7\right) }{\sqrt{2}}\frac{1}{\sqrt{m\delta }}+
\frac{1}{4\sqrt{2}}                                              \label{A1:VIa}
\\
&&\times \left( \frac{\sqrt{3m}}{m^{2}-\omega _{\text{tn} }^{2}}+\frac{\cosh
^{-1}\left( 7\right) }{m^{3/2}}\right) \sqrt{\delta }+O\left( \delta ^{\frac{
3}{2}}\right). \nonumber
\end{eqnarray}
Using Eqs.~(\ref{A1:V}) and (\ref{A1:VI}),  we  obtain the dependence of $E$ on
$Q$ in the thick-wall regime:
\begin{equation}
E = m \left\vert Q \right\vert - \frac{h}{3!}\frac{m^{2}-
\omega _{\text{tn} }^{2}}{12m^{3}} \left\vert Q \right\vert^{3}+
O\left(\left\vert Q \right\vert^{5}\right).                      \label{A1:VII}
\end{equation}
From  Eqs.~(\ref{A1:V}) -- (\ref{A1:VIa}), it follows  that in  the  thick-wall
regime, the  soliton's  Noether  charge  and  energy vanish as $\sqrt{\delta}$,
whereas the soliton's effective size diverges as $1/\sqrt{\delta}$.

In the thin-wall regime, the squared phase frequency tends to its minimum value:
$\omega^{2} \rightarrow \omega_{\text{tn}}^{2}$.
In this  regime,  the  Noether charge, the energy, and the width at half-height
of the one-dimensional Q-ball behave as follows:
\begin{eqnarray}
Q &=&\text{sgn}\left(\omega\right)\sqrt{\frac{3}{h}}
\omega_{\text{tn}}\left[ \ln \left( \frac{2\left(m^{2}-
\omega_{\text{tn}}^{2}\right)}{\omega_{\text{tn}}\bar{\delta} }\right)
\right.  \nonumber \\
&&\left. -\sqrt{\frac{2\omega_{\text{tn}}\bar{\delta} }
{m^{2} - \omega_{\text{tn}}^{2}}} + O\left(\bar{\delta}\right)\right],
                                                                \label{A1:VIII}
\end{eqnarray}

\begin{eqnarray}
E &=&\sqrt{\frac{3}{h}}\text{$\omega $}_{\text{tn}}^{2}\left[ \ln \left(\frac{
2\left( m^{2}-\text{$\omega $}_{\text{tn} }^{2}\right) }
{\text{$\omega$}_{\text{tn}}\bar{\delta} }\right) \right.         \label{A1:IX}
\\
&&\left. + \frac{m^{2}}{\text{$\omega $}_{\text{tn} }^{2}} - 1 -
\sqrt{\frac{2\text{$\omega $}_{\text{tn} }\bar{\delta}}{m^{2} -
\text{$\omega $}_{\text{tn}}^{2}}}+O\left(\bar{\delta}\right)\right], \nonumber
\end{eqnarray}
and
\begin{eqnarray}
L &=&2^{-1}\left( m^{2}-\omega _{\text{tn}}^{2}\right) ^{-1/2}\ln \left( \frac{
18\left( m^{2}-\omega _{\text{tn}}^{2}\right) }
{\omega_{\text{tn}} \bar{\delta}  }\right)  \nonumber \\
&&+4\frac{\sqrt{2}}{3}\frac{\sqrt{\omega _{\text{tn}} \bar{\delta}}}
{m^{2} - \omega _{\text{tn}}^{2}}+O\left( \bar{\delta} \right),
                                                                 \label{A1:IXa}
\end{eqnarray}
where  the  variable  $\bar{\delta}  =  \left\vert\omega \right\vert - \omega_{
\text{tn}}$.
From Eqs.~(\ref{A1:VIII}) and (\ref{A1:IX}), we obtain the dependence of $E$ on
$Q$ in the thin-wall regime:
\begin{eqnarray}
E &=&\omega_{\text{tn}} \left\vert Q \right\vert  +
\sqrt{\frac{3}{h}}\left( m^{2} - \omega _{\text{tn}}^{2}\right)    \nonumber
\\
&& + O\left( \exp \left( - \sqrt{\frac{h}{3}}\frac{\left\vert Q \right\vert}
{\omega _{\text{tn}}}\right) \right).                              \label{A1:X}
\end{eqnarray}
From Eqs.~(\ref{A1:VIII}), (\ref{A1:IX}),  and  (\ref{A1:IXa}), it follows that
the Noether charge, the energy, and  the  effective size of the one-dimensional
Q-ball logarithmically diverge in the  thin-wall regime.

\bibliography{article}

\end{document}